\documentclass[12pt]{article}
\begin{document}
\baselineskip 18pt
\newcommand{\Tr}{\mbox{Tr\,}}
\newcommand{\Dirac}{/\!\!\!\!D}
\newcommand{\beq}{\begin{equation}}
\newcommand{\eeq}[1]{\label{#1}\end{equation}}
\newcommand{\bea}{\begin{eqnarray}}
\newcommand{\eea}[1]{\label{#1}\end{eqnarray}}
\renewcommand{\Re}{\mbox{Re}\,}
\renewcommand{\Im}{\mbox{Im}\,}
\begin{titlepage}
\hfill  CERN-TH/98-330 NYU-TH/98/10/02 hep-th/9810063
\begin{center}
\hfill
\vskip .4in
{\large\bf N=6 Supergravity on $AdS_5$ and the $SU(2,2/3)$ Superconformal
Correspondence}
\end{center}
\vskip .4in
\begin{center}
{\large S. Ferrara$^a$, M. Porrati$^{a,b,c}$ and
A. Zaffaroni$^a$\footnotemark}
\footnotetext{e-mail: sergio.ferrara@cern.ch,
massimo.porrati@nyu.edu, alberto.zaffaroni@cern.ch}
\vskip .1in
(a){\em Theory division CERN, Ch 1211 Geneva 23, Switzerland}
\vskip .1in
(b){\em Department of Physics, NYU, 4 Washington Pl.,
New York, NY 10003, USA\footnotemark}
\footnotetext{Permanent Address}
\vskip .1in
(c){\em Rockefeller University, New York, NY
10021-6399, USA}
\vskip .1in
\end{center}
\vskip .4in
\begin{center} {\bf ABSTRACT} \end{center}
\begin{quotation}
\noindent
It is argued that N=6 supergravity on $AdS_5$, with gauge group
$SU(3)\times U(1)$ corresponds, at the classical level, to a subsector of
the ``chiral'' primary operators of N=4 Yang-Mills theories. This projection
involves a ``duality transformation'' of N=4 Yang-Mills
theory and therefore can be valid if the coupling is at a self-dual point,
or for those amplitudes that do not depend on the coupling constant.
\end{quotation}
\vfill
CERN-TH/98-330 \\
October 1998
\end{titlepage}
\eject
\noindent
\section{Introduction}
The recent understanding of many features of the $AdS_{d+1}^{2N}/CFT_d^N$
correspondence \cite{malda,pol,witten}, where $d$ is the dimension of the
boundary conformal field
theory and $N$ the number of (boundary) Poincar\'e supersymmetries, naturally
points to investigate more general supergravity theories in $AdS^N_d$, that
have lower supersymmetry, as well
the cases when these theories have no obvious
interpretation in terms of standard compactifications.
Among the latter is the class of theories for which the number of
supersymmetries $N$ is not a power of two, and that can not, therefore,
be obtained by standard compactifications of superstring theories.

The most familiar examples are  $O(N)$ $AdS_4$ supergravities with $N=5,6$,
corresponding to three-dimensional superconformal algebras $OSp(N/4)$
\cite{nahm}.

Another example, which is the one considered in this note, is $N=6$
$AdS_5$ supergravity\footnote{Stringy constructions of $N=6$ 5D supergravity in
flat space $M_5$ were given in \cite{koun,harv}} in five dimensions, associated
to the $N=3$ superconformal algebra in $d=4$ dimensions.

An immediate puzzling feature of this particular case is, of course, the fact
that $N=3$, $d=4$ Yang-Mills theory is known to be the same as $N=4$
Yang-Mills theory, although the corresponding superalgebras ($SU(2,2/N)$
\cite{nahm})
are different. Multiplets with spin greater than one are however different in
the two theories, and $N=6$ supergravity is not the same as $N=8$
\cite{cremmer}.

In this note, we show that the existence of $N=6$ supergravity, at least
classically, may be related to some properties of the OPE of $N=4$ Yang-Mills
theory, at least in a regime in which a certain symmetry is supposed to hold.
This symmetry truncates the $N=4$ ``chiral'' operators to a
subset of $N=3$ operators, which do not contain the additional $AdS_5$
representations which complete $N=6$ to $N=8$, $AdS_5$ supergravity.
\section{$N=6$ Supergravity and its Symmetries}
In their several papers on supergravity on AdS$_5$, G\"unaydin, Romans and
Warner discussed $N=6$ supergravity \cite{gun} as a consistent truncation of
$N=8$
supergravity \cite{n=8}.
The two $N=6$ multiplets are the graviton multiplet (containing three complex
gravitinos), and the gravitino multiplet. The components of these multiplets
are given in table 3 in~\cite{gun}, together with their quantum number under
the $SU(3)\times U_D(1)$ subgroup of the $USp(6)$ hidden symmetry of the $N=6$
theory. The $N=8$ graviton multiplet decomposes in an $N=6$ graviton  and an
$N=6$ gravitino multiplet. $N=6$ supergravity is obtained by consistently
truncating the $N=8$ supergravity to the $N=6$ graviton multiplet.
We notice that, in this truncation, the original gauge group $SU(4)$ is
broken to $SU(3)\times U(1)$, and that the $N=8$ dilaton field, belonging
to the $N=6$ gravitino multiplet, disappears from the spectrum.

A crucial ingredient in the truncation is the fact that in the original $N=8$
theory there is a $U(1)$ symmetry commuting with the gauge group $SU(4)$. This
comes from the fact that \cite{gun}
\beq
E_{6(6)}\rightarrow SL(2;R)\times SL(6;R)\rightarrow U_S(1)\times SU(4).
\eeq{dddddd}
On the other hand, the maximal compact group $USp(8)$ in $E_{6(6)}$ can be
decomposed in two different ways down to $SU(3)\times U(1)$:
\beq
{\rm I}:\qquad\qquad USp(8)\rightarrow SU(4)\times U_S(1)\rightarrow
SU(3)\times U_R(1)\times U_S(1).
\eeq{I}
Here $U_S(1)$ is a subgroup of the $SL(2;R)$ symmetry of the theory
whose $SL(2;Z)$ subgroup can be identified with the S-duality group of both
the underlying type IIB theory and the N=4 boundary Yang-Mills theory.
\beq
{\rm II}: \qquad\qquad USp(8)\rightarrow USp(6)\times SU(2)\rightarrow SU(3)
\times U_D(1)\times U_1(1).
\eeq{II}
Under $USp(6)$, the 15 original ungauged vectors transform as 14+1, and the
$SU(3)\times U(1)$ gauge bosons come from the octect in the
$14\rightarrow (8,0)+(3,1)+(3,-1)$ decomposition of
$USp(6)\rightarrow SU(3)\times U_D(1)$, as well as
from the $USp(6)$ singlet in the original decomposition of
$USp(8)\rightarrow USp(6)\times SU(2), 27\rightarrow (14,1)+(6,2)+(1,1)$.

Clearly, there are only two independent $U(1)$ factors: the pair of $U(1)$s in
the decomposition II are linear combinations of the $U(1)$s in the
decomposition I, and viceversa.
The $U(1)$s in decomposition I correspond to symmetries of the original $N=4$
Yang-Mills theory, or type IIB theory on $AdS_5\times S^5$, namely, a $U(1)$
subgroup of the R-symmetry $SU(4)$ and a (discretized) subgroup of $SL(2;Z)$.
The $U(1)$s in decomposition II are more suitable for discussing the
structure of N=6 representations and the consistent truncation from $N=8$,
as we shall see shortly.

The $U_S(1)$ transformation is an automorphism of the $SU(2,2/4)$ superalgebra,
acting as $e^{i{3\over 4}\alpha}Q_L,e^{-i{3\over 4}\alpha}S_L$ on Poincar\'e
and conformal left supercharges \cite{haag}, respectively. It corresponds to a
$\gamma_7$ transformation in the $O(4,2)$ covariant formulation
\cite{mack,ferrara}.
\section{The $N=3$ Truncation}
The very fact that the $N=8$ supergravity admits a consistent truncation to
$N=6$ seems to suggest that, in a certain regime, we can define a closed $N=3$
subalgebra of gauge singlet operators in $N=4$ Yang-Mills theories.

In complete analogy with the $N=4$ case \cite{FFZ}, we can construct the $N=3$
``chiral'' spectrum of operators (or, equivalently, $N=6$ KK excitations) by
considering tensor products of the fundamental $N=3$ singleton representation
that can be obtained by considering the $N=3$ pure Yang-Mills theory in $N=3$
superspace.

Following~\cite{stelle}, such singleton representation is described in $N=3$
superspace by a superfield-strength, which is a Lorentz scalar and an $SU(3)$
triplet, $W_i(x,\theta)$. This $W_i$ satisfies some constraints, which can be
found in~\cite{stelle}, however its physical components lie in just the
first few terms of the $\theta$ expansion,
\beq
W_i(x,\theta)=\phi_i+\theta_{iL}\lambda_L +\theta^l_R\lambda_R^m\epsilon_{lmi}
+\theta^l_R\theta^m_R\epsilon_{lmi}F_R^+ + ...\qquad .
\eeq{super}
Notice that both $\theta_L$ and $\theta_R$ contribute to the physical
components of $W_i$. Under $N=1$ supersymmetry, $W_i$ decomposes into three
``chiral'' multiplets containing the physical fields $(\phi_i, \lambda_i)$ and
a
vector multiplet containing $(F,\lambda)$.

Let us describe the $U_D$ and $U_1$ quantum numbers of the $W_i$ components.
We see from eq.~(\ref{II}) that $U_1$ commutes with $USp(6)$ while $U_D$ must
act as a R-symmetry.
We can assign $U_D(1)$ charge $-1/2$ to $\phi_i$ and +1 to the $\theta$s, and
$U_1(1)$ charge +3/2 to all the components of the multiplet.
The transformation rules:
\bea
U_D(1)&:\qquad\qquad W_i(\theta)\rightarrow e^{-{1\over 2}i\alpha}
W_i(e^{i\alpha}\theta),\\
U_1(1)&:\qquad\qquad W_i(\theta)\rightarrow e^{{3\over 2}i\alpha}W_i(\theta),
\eea{simm}
give for the components:
\bea
U_D(1)&:\qquad\qquad \phi_i \left (-{1\over 2}\right ), \lambda_{iL}
\left (-{1\over 2}\right ), F^-_L \left (-{3\over 2}\right ),
\lambda_L \left (-{3\over 2}\right ),\\
U_1(1)&:\qquad\qquad \phi_i \left ({3\over 2}\right ), \lambda_{iL}
\left (-{3\over 2}\right ), F^-_L \left (-{3\over 2}\right ), \lambda_L
\left ({3\over 2}\right ).
\eea{trasf}
Notice that since the above symmetries act on $F_L$ as a duality rotation,
they can only be realized in free-field theory as continuous invariances, and
in non-perturbative Yang-Mills theory at the self-dual point as discrete
subgroups. Notice also that  $U_D$ commutes with $N=1$ supersymmetry.

We give also the quantum numbers of the gauge group $U_R(1)$.
They can be easily obtained by decomposing
$SU(4)\rightarrow SU(3)\times U_R(1)$,
\beq
U_R(1):\qquad\qquad \phi_i (-2), \lambda_{iL}
(1), F^-_L \left ( 0 \right ), \lambda_L
(-3).
\eeq{gauge}
This symmetry is known to be a continuous symmetry of perturbative Yang-Mills
theory.

As for the $U_S(1)$ symmetry, that commutes with $SU(4)$ and corresponds to
the linearly realized $U(1)$ subgroup of $SL(2;R)$, we can define it directly
in terms of the $N=4$ superfield:
\beq
W_{[AB]}\rightarrow W_{[AB]}(e^{i{3\over 4}\alpha}\theta_a), \qquad A,a=1,..,4.
\eeq{NNNN}
This gives for the components:
\beq
U_S(1)=U_D(1)-{1\over 4}U_R(1):\qquad\qquad \phi_i \left ( 0\right ),
\lambda_{iL},\lambda_L
\left (-{3\over 4}\right ), F^-_L \left (-{3\over 2}\right ).
\eeq{SL}

It is clear that several relations among the various $U(1)$s hold, since
only two of them are independent. For example, we have
\beq
U_1(1)= U_D(1)-U_R(1)=-{3\over 4}U_R(1) +U_S(1).
\eeq{ddd}

It is crucial for us that there exists $U_1(1)$, which commutes with $USp(6)$.
We can use this $U(1)$ to define the truncation of the $N=8$ supergravity
theory, or, equivalently, the truncation of the $N=4$ Yang-Mills theory.
We can express $U_1=-{3\over 4}U_R+U_S$ as a linear combination of a $U(1)$
subroup of the $SU(4)$ R-symmetry of $N=4$ SYM and a (discretized) $U(1)$
subgroup of the S-duality group $SL(2;Z)$. By choosing an appropriate element
of $SL(2;Z)$, the discretized $U_1$ is a (non-perturbative) symmetry of the
$N=4$ SYM theory at the self-dual point.

The two spin-2 and spin-3/2 $N=6$ multiplets, which will be denoted $G$ and
$g$, respectively, can be written as bilinears of the singleton fields:
\bea
&G_i^j=\Tr\left (W_i\bar W^j -{1\over 3}\delta_i^jW_k\bar W^k\right )\\
&g_{ij}=\Tr\left (W_iW_j\right ).
\eea{def}
The $\theta$ expansion of these superfields exactly reproduces the structure
and $SU(3)\times U_D(1)$ quantum number of the graviton and gravitino $N=6$
multiplet, as shown in table 3 of~\cite{gun}.

At the bilinear level, the other multiplet, corresponding to the radial mode,
is the Konishi  multiplet  $\Tr W_i\bar W^i$ ,
which is not a ``chiral'' operator \cite{proc}.

It is obvious that, under $U_1$, the two superfields transform as,
\bea
&G_i^j\rightarrow G_i^j\\
&g_{ij}\rightarrow e^{3i\alpha}g_{ij}
\eea{trunc}
We see that, as promised, $U_1$ can be used to eliminate the unwanted
gravitino multiplet and to truncate $N=8$ supergravity to $N=6$. Note that the
Konishi multiplet is not projected out by the truncation as required by
consistence of the OPE of two stress-energy tensors in Yang-Mills
theory~\cite{anselmi}.

This analysis can be extended to the whole set of $N=4$ ``chiral'' operators.
We can use the $U_1$ projection to define an $N=3$ subset of ``chiral''
operators.
For $N$-extended supersymmetry, a long-multiplet has maximum spin $(N/2,N/2)$.
Therefore, long multiplets of $N=6$ have maximum spin at least equal to
3. It then follows that the ``chiral'' $N=4$ primary fields, having at most
spin
2, are also (reducible) short multiplets of the $SU(2,2/3)$ superalgebra.
So, the quantization of the spectrum is, also in the $N=3$ case,
a consequence of supersymmetry.

The $N=4$ ``chiral'' operators --which can be written as $\Tr W^p$~\cite{FA},
where $W$ is the $N=4$ singleton, as defined in harmonic
superspace~\cite{west}-- decompose in $N=3$ operators that are products of
$W$ and
$\bar W$, with suitable symmetrizations and removal of traces.
The $U_1$ projection eliminates all the products that do not involve an
equal number of $W$ and $\bar W$. The
expected $N=3$ ``chiral'' operator has, therefore, the general form
\beq
W^{2p} = \Tr (W_{i_1}\bar W_{i_2}... W_{i_{2p-1}}\bar W_{i_{2p}}).
\eeq{boh}
Of course, the $U_1$ projection involving an element of $SL(2;Z)$
is discrete.
This implies that particular powers of $W$ and $\bar W$ may survive the
projection. If the charge of $W_i$ is $2\pi /k$, strings of operators of the
form $W_{1_1}....W_{i_k}$ or  $\bar W_{1_1}....\bar W_{i_k}$ are allowed in
eq. ~(\ref{boh}).
\section{The Truncation on the Supergravity Side.}
The argument in the previous section can be retrieved by reasoning in the
$N=8$ supergravity context.

$N=8$ supergravity on $AdS_5$ can be regarded as the supersymmetric completion
of a gauged $\sigma$-model with $G/H=E_{6(6)}/USp(8)$ and gauge group $SU(4)$.

The relevant decomposition to obtaining $N=6$ supergravity is
\beq
E_{6(6)}\rightarrow SU^*(6)\times SU(2),
\eeq{jjjjjjjj}
with the following embedding of the previous defined $U(1)$'s: $U_1(1)\subset
SU(2)$ and $U_D(1)\subset SU(3)\in SU^*(6)$. The truncation defining $N=6$
gauged supergravity is obtained by retaining only $U_1(1)$ singlets. This leads
to the
identification of $U_D(1)$ with $U_R(1)$ on the singlet modes, according to
eq.~(\ref{ddd}). As a result, $G/H=E_{6(6)}/USp(8)$ with gauge group $SU(4)$
is truncated to $SU^*(6)/USp(6)$\footnote{Note that the rank 6 coset
$E_{6(6)}/USp(8)$, as a solvable Lie algebra \cite{dauria}, decomposes as
\beq
{\rm Solv}(E_{6(6)}/USp(8))={\rm Solv} (SU^*(6)/USp(6))+{\rm
Solv}(F_{4(4)}/USp(6)\times SU(2))
\eeq{nbv}
where the rank 2 and rank 4 cosets above correspond to the following
decomposition $42=(14,1)+(14',2)$ of the original $N=8$ scalars with respect to
$USp(6)\times SU(2)$.}
 with gauge group $SU(3)\times U(1)$ \cite{gun}.

The extension of this argument to the massive states requires that one can
define the $N=6$ truncation directly
in 10 dimensions. No {\em geometrical} symmetry can be used to truncate
$N=8$ to $N=6$ on $AdS_5\times S^5$, but a combination of isometries and
a duality transformation preserves the right number of Killing spinors.
This is easily seen by recalling that the Killing spinor transforms as
a 4 of $SU(4)$, the isometry group of $S^5=SO(6)/SO(5)\approx SU(4)/USp(4)
$~\cite{pvn}. Using the harmonic expansion on $S^5$, the Killing spinor
reads~\cite{salam}
\beq
\epsilon_i(x)=D_{ij}^4[L^{-1}(x)]c_j,\;\;\; x\in S^5.
\eeq{salam}
Here $D^4(g)$ is the $4\times 4$ matrix transforming in the 4 of $SU(4)$;
$L(x)$ is the coset representative of $x$ in $SU(4)$, and $c_j$ are arbitrary
constants. Isometries of $S^5$ act on $D^4$ as right multiplications, while
the linearly-realized $U_S(1)$ in $SL(2,R)$ acts as multiplication by a phase:
\bea
g\in SU(4)&:& D_{ij}^4[L^{-1}(x)] \rightarrow D_{il}^4[L^{-1}(x)]
D_{lj}^4[g^{-1}],\nonumber \\
h\in U(1)_S&:& D_{ij}^4[L^{-1}(x)] \rightarrow \exp(i\theta)
D_{ij}[L^{-1}(x)].
\eea{iso}
By choosing a projection acting as an $SU(4)$ isometry
$g=\exp(i\theta\lambda)$, $\lambda={\rm diag}\,(1,1,1,-3)$, combined with a
$U_S(1)$
$h=\exp(i\theta)$, one can project away the Killing spinors with $c_4\neq 0$,
and preserve only 6 of the original 8 supersymmetries of the background.

\section{Hidden Symmetry in $N=4$ Yang-Mills Theory}
Let us examine the consequence of the truncation defined in the previous
section on the Green functions of the $N=4$ Yang-Mills theory.

The vanishing of amplitudes with unequal number of $W$ and $\bar W$ $N=3$
field strengths implies a certain set of relations among $N=4$ correlators.

The $N=3$ correlators with only one gravitino, $\langle G_1...G_ng\rangle $,
vanish,
as any amplitude with an odd number of $g$. However, amplitudes of the form
$\langle G_1...G_ng_1...g_m\bar g_1...\bar g_m\rangle $ may not vanish.
As an example,
in the decomposition of $SU(4)$ into $SU(3)$, the scalars in the $N=4$
graviton multiplet decompose as
\beq
20_R\rightarrow 8+6+\bar 6,\qquad 10\rightarrow 1+3+6,\qquad 1\rightarrow 1.
\eeq{JJJJ}
The $N=3$ supergravity only contains the 14 scalars in $8+3+\bar 3$.
Therefore, the dilaton-axion belongs to $g=\Tr W^2$ and (partially) decouples
from $N=3$ amplitudes.

The non-perturbative projection $U_1$ is in general discrete. If it acts as a
$\pi$ phase on the gravitino $g$, then the relation $\langle G_1...G_mg\rangle
=0$ is
still valid, but
$\langle g^2\rangle $, for example, is allowed as a composite operators,
at the non-perturbative level. In the sequence
$\langle \Tr W^{4n}W\bar W ... W\bar W\rangle $ this gives extra allowed
``chiral'' operators
which show up only at the $p=4$ level.

In the free-field theory, however, the $U_1(1)$ symmetry
$W_i\rightarrow e^{i\beta}W_i$ can be used to give a
stronger selection rule: all correlators with an odd number of $W_i$ must
vanish. Let us analyse the consequence of this rule for the
free-field theory, and, therefore, for all amplitudes that do not depend
on the coupling constant.
Composites like $\Tr W^2$ can only have non-vanishing even n-point functions.
Moreover, OPEs involving $\Tr W\bar W \Tr W\bar W$ can never produce
$\Tr W^2$ operators, since the three-point function $\langle \Tr W\bar W \Tr
W\bar W
\Tr W^2\rangle $ vanishes.
In the $N=3$ language, the dilaton is a high component of
the superfield $\Tr W_iW_j$.
The dilaton 3-point function vanishes,
but the four-point function does not. The dilaton has no mixed 3-point
functions $\langle \Tr W_iW_j\Tr W_l\bar W_k \Tr W_m\bar W_p\rangle $.
It should be noticed that this symmetry is violated by perturbative
corrections. It may also be a symmetry at the self-dual coupling, as the
consistent
truncation of the $N=8$ supergravity seems to indicate, up to corrections  of
order $O(1/{\rm N})$.



We want to check these selection rules directly in the CFT.
In free-field theory, a consistent subset of ``chiral'' operators is the
irreducible $SU(3)$ components of $\Tr (W_{i_1}\bar W_{i_2}... W_{i_{2p-1}}
\bar W_{i_{2p}})$.
Using $N=2$ techniques, it can be shown that any correlation function with
different number of $W$ and $\bar W$ does in fact vanish. This because the
$N=2$ hypermultiplet in $W$ has vanishing $\langle \phi\phi \rangle $
propagator, but not
vanishing $\langle \phi\bar\phi \rangle $ propagator.
This vanishing is not surprising because $U_1$ is a symmetry of the free-field
theory.
We conclude that, in free-field theory and for those amplitudes that are
independent of the coupling constant, the selection rules are valid. In the
fully interacting theory, however, only a discrete subgroup of $U_1$ can be a
symmetry, and only at a self-dual
value of the coupling constant. This suggest that the selection rules may be
valid for $N=4$ Yang-Mills at the self-dual point.

We can also use the CFT/AdS correspondence to give further support to the
validity of the selection rules at the self-dual point.
We can start with $N=4$
SYM and its dual description as type IIB string theory on $AdS_5\times S^5$
at the self-dual value of the coupling, and perform a projection with the
symmetry $U_1$, which is a combination of an element of $SL(2;Z)$
and a discrete element of the isometry group of $S^5$. This consistently
truncates to an $N=6$ theory, which has
the $N=6$ gauged supergravity as the effective action for the massless modes.
Since the coupling constant is fixed, it is clear that we are not exploring
the t'Hooft large-N limit. Maldacena's conjecture, however, applies whenever
we can trust the supergravity approximation, i.e. whenever we can ignore
higher-dimension operators in the expansion of the effective action of type
IIB superstrings.
One such limit is the large-N limit at fixed string coupling constant (see
for instance~\cite{banks}). In this limit, the $\alpha'$
expansion corresponds to the 1/N expansion of the theory. We conclude that the
supergravity description supports the existence of selection rules for the
$N=4$ theory at the self-dual point at least in the large-N limit.

The selection rules clearly extend to arbitrary value of the coupling constant
for all amplitudes that do not depend
on the coupling constant in the large-N limit. According to a conjecture in
\cite{seiberg}, all three point functions of ``chiral'' operators belong to
this class of amplitudes, and the result is valid also for finite
N. A perturbative computation that agrees with this conjecture was
performed in \cite{doker,sok}.
It is known instead that the four-point functions
of ``chiral'' operators do depend on the coupling constant
\cite{banks}\footnote{Partial
results on the four-point functions, based on  the $CFT/AdS$
correspondence, can be found in the literature \cite{four}.}.


{\bf Acknowledgements}\vskip .1in
\noindent
We would like to thank D. Anselmi, R. D'Auria, P. Fr\'e, Y. Oz, M. Trigiante,
and especially E. Sokatchev for discussions. M.P.
is supported in part by NSF grant no. PHY-9722083.
S.F. is supported in part by the EEC under TMR contract ERBFMRX-CT96-0090, ECC
Science Program SCI$^*$ -CI92-0789 (INFN-Frascati), DOE grant
DE-FG03-91ER40662.

\end{document}